\documentclass[12pt]{article}
\usepackage{amssymb,amsmath,latexsym}
\title{Quantization of edge currents  \\ for continuous magnetic operators}

\author{J. Kellendonk$^{1}$,
H. Schulz-Baldes$^{2}$,
\\
\\
\\
$^1$ {\small School of Mathematics, Cardiff University, 
Cardiff, CF24 4YH, Wales}
\\
$^2$ {\small Fachbereich Mathematik, TU Berlin, 
Strasse des 17. Juni 136, 10623 Berlin, Germany}
}

\date{ }

\newtheorem{thm}{Theorem}

\newtheorem{prop}{Proposition}
\newtheorem{lem}{Lemma}
\newtheorem{cor}{Corollary}

\newcommand{\Real}{\mathbb R}
\newcommand{\RR}{\Real}

\newcommand{\Complex}{\mathbb C}
\newcommand{\CC}{\Complex}

\newcommand{\Ta}{{\mathcal{A}}}
\newcommand{\Aa}{\mathcal{A}_\infty}

\newcommand{\Uu}{{\cal U}}

\newcommand{\ZZ}{{\bf Z}}

\newcommand{\PP}{{\bf P}}

\newcommand{\EE}{{\bf E}}

\newcommand{\Tr}{\mbox{\bf Tr}}  
\newcommand{\TV}{{\mathcal T}}
\newcommand{\TVh}{\hat{{\mathcal T}}}

\def\esssup{\mathop{\rm ess\,sup}}
\newcommand{\n}{{\vec n}}

\newcommand{\bew}{{\bf Proof:}}
\newcommand{\eb}{\hfill $\Box$}

\newcommand{\om}{\omega}
\newcommand{\oh}{{\hat{\omega}}}
\newcommand{\x}{{\vec x}}
\newcommand{\y}{{\vec y}}

\newcommand{\hull}{\Omega}

\begin{document}

\maketitle

\begin{abstract}
For a magnetic Hamiltonian on a half-plane given as the sum of the
Landau operator with Dirichlet boundary conditions and a random
potential, a quantization theorem for the edge currents is proven.
This shows that the concept of edge channels also makes sense in
presence of disorder. Moreover, Gaussian bounds on the
heat kernel and its covariant derivatives are obtained.
\end{abstract}


\vspace{1cm}

\section{Introduction}

Topological quantization of edge currents has been proven rigorously
only for discrete magnetic Schr\"odinger operators \cite{SKR,KRS,EG}.
The purpose of this work is prove similar results also for continuous
Schr\"odinger operators. In order to describe the main result, 
$H_L$ denotes the Landau operator on $L^2(\RR^2)$ and 
$V$ is a differentiable potential given as sum of a periodic and a
random part. Let
$\hat{H}$ denote the restriction of $H_L+V$
to the half-plane with Dirichlet boundary conditions and let
$J_1=\imath[\hat{H},X_1]$ be the current operator along the
boundary. Suppose that the interval $\Delta$ is a gap of 
$H_L+V$ (but this is then {\it not} a gap of $\hat{H}$) and
that $G:\RR\to[0,1]$ is a decreasing differentiable function equal
to $1$ to the left of $\Delta$ and $0$ to its right. Hence its
derivative $G'$ is
negative and supported by $\Delta$. Furthermore let $\chi$ be a
smooth, positive, and compactly supported function on $\RR$ with unit
integral, namely $\int \chi =1$. Under these conditions, it is shown 
that $\chi(X_1)J_1G'(\hat{H})\chi(X_1)$ is a traceclass operator
and that 

$$
\EE\;\Tr
\left(
\chi(X_1)J_1G'(\hat{H})\chi(X_1)
\right)
\;=\;\frac{1}{2\pi}\;
\mbox{Ind}\mbox{ , }
$$

\noindent where $\EE$ denotes the disorder average and
$\mbox{Ind}\in\ZZ$ is the index of a certain Fredholm operator which depends
on $\Delta$, but not on the choice of the functions $G$ and $\chi$. 
The index is also equal to the non-commutative
winding number of the unitary operator
$\Uu=\exp(-2\pi\imath\,G(\hat{H}))$. 

\vspace{.2cm}

The quantity
$\EE\;\Tr\left(\chi(X_1)J_1G'(\hat{H})\chi(X_1)\right)$ can
physically be interpreted as the conductivity of the edge. 
Indeed, $-G'(\hat{H})\geq 0$ is a density matrix of edge states in
$\Delta$ which is normalized because $-\int G'=1$. 
The operator $\chi(X_1)J_1G'(\hat{H})\chi(X_1)$ gives the
corresponding current along the boundary and within a
in strip of unit width which is
perpendicular to the boundary. According to the above, 
its averaged trace is quantized.
Another interpretation, fully developed in \cite{SKR}, is obtained
when the smooth function $G'(\hat{H})$ approximates $\frac{1}{|\Delta'|}
\chi_{\Delta'}(\hat{H})$ where $\chi_{\Delta'}$ is the indicator
function of the interval $\Delta'\subset \Delta$. The boundaries of
$\Delta'$ are thought to be the local Fermi levels at the upper and
lower boundary of a bar-like sample. Then the above quantity measures
the net edge current, namely the sum of the current along the upper
boundary and the (reversed) one on the lower boundary. This net edge
current is hence quantized. Moreover the above index is equal to
the bulk conductivity as given by the Kubo formula as long as the Fermi
level lies in $\Delta$ 
(the proof of this fact is deferred to a forthcoming work \cite{KS}).
Hence both edge and bulk currents in the bar are
quantized with the same Fredholm index, a fact of crucial importance
for the quantum Hall effect (see \cite{SKR,KRS} and references
therein). 

\vspace{.2cm}

Let us briefly discuss the hypothesis. The gap condition on (the
density of states of) $H$ should
be satisfied in clean high mobility samples. However, the quantization 
of the edge currents probably also holds under a weaker dynamical
localization condition, just as the quantization of the Kubo-Chern
formula does \cite{BES,AG}. The Dirichlet boundary conditions could also
be replaced by a soft edge modeled by a confining potential.
Technical modifications would mainly be needed in Section
\ref{sec-lifting}. 

\vspace{.2cm}

Compared to \cite{KRS}, the main technical difficulties en route
concern proving Gaussian bounds on the 
heat kernels as well as their (covariant) derivatives. Even though
this is known for the heat kernels themselves \cite{Sim82,Dav,BHL}, 
to our knowledge the
derivatives have not been controlled as explicitely. 
The proof of the quantization of edge currents 
is based on an
index theorem for covariant families of unitaries (Section
\ref{sec-winding}).

\section{Magnetic Hamiltonians}
\label{sec-Ham}

Let ${\vec X}=(X_1,X_2)$ be the position operator on $L^2(\RR^2)$ and 
${\vec \partial}=(\partial_1,\partial_2)$ the associated partial
derivatives. Then ${\vec X}$ and $\imath {\vec \partial}$ 
are self-adjoint with common core
$C^\infty_c(\RR^2)$, the smooth functions with compact support. 
Setting $\gamma=\frac{qB}{\hbar}$, 
the Landau operator in the Landau gauge is then
given by

$$
H_L
\;=\;
\frac{\hbar^2}{2m}\left(\imath\partial_1
-\gamma\,X_2\right)^2 + \frac{\hbar^2}{2m}
\left(\imath\partial_2\right)^2
\mbox{ . }
$$

\noindent In order to simplify notations, units are chosen such
that $\hbar^2/m =1$.
With the following operators (all with common core 
$C^\infty_c(\RR^2)$)

$$
D_1\;=\;\imath\partial_1-\gamma X_2
\mbox{ , }
\qquad
D_2\;=\;\imath\partial_2
\mbox{ , }
\qquad
K_1\;=\;\imath\partial_1
\mbox{ , }
\qquad
K_2\;=\;\imath\partial_2-\gamma X_1
\mbox{ , }
$$

\noindent it can be written as $H_L=\frac{1}{2}(D_1^2+D_2^2)$. 
One readily verifies that 
both $D_{1,2}$ commute with both $K_{1,2}$. Hence the Landau
operator has a large symmetry group and its spectrum is infinitely
degenerated. The $D_{1,2}$ and $K_{1,2}$ are
respectively the generators of the Landau translations $T({\vec
\xi}\,)$ and 
the magnetic translation operators $U({\vec \xi}\,)$ 
defined by (for
$\psi\in L^2(\RR^2)$ and $\x=(x_1,x_2)\in\RR^2$)

%

\begin{eqnarray*}
(U({\vec \xi}\,)\psi)(\x\,)
& = &
\hat{\Phi}({\vec \xi},\x-{\vec \xi}\,)\,\psi(\x-{\vec \xi}\,)
\mbox{ , }
\qquad
\hat{\Phi}({\vec \xi},\x\,)\;=\;e^{-\imath\gamma\xi_2x_1}
\mbox{ , }\\
(T({\vec \xi}\,)\psi)(\x\,)
& = &
{\Phi}({\vec \xi},\x-{\vec \xi}\,)\,\psi(\x-{\vec \xi}\,)
\mbox{ , }
\qquad
{\Phi}({\vec \xi},\x\,)\;=\;
e^{-\imath\gamma\xi_1x_2}
\mbox{ . }
\end{eqnarray*}

\noindent It can be easily verified that 
the following relations hold

$$
U({\vec \xi}\,)U(\vec{\eta}\,) \;=\;\hat{\Phi}({\vec \xi},{\vec \eta}\,)
U({\vec \xi}+{\vec \eta}\,)
\mbox{ , }
\qquad
T({\vec \xi}\,)T({\vec \eta}\,) 
\;=\;
{\Phi}({\vec \xi},\vec{\eta}\,)T(\vec{\xi}+\vec{\eta}\,)
\mbox{ . }
$$

\vspace{.2cm}

The aim is now to add a potential to $H_L$ having possibly a periodic and
a disordered component. Let the set $\hull$ of configurations of the
potential be compact and let  
$\RR^2$ act homeomorphically on it.
This action will simply be denoted by $\om\mapsto \x\cdot\om$,
$\omega\in\Omega$. The set $\Omega$ is often called the hull.
Furthermore let $\PP$ be an invariant and ergodic
probability measure on the hull. 
Given a measurable positive function $V\in L^\infty(\Omega,\PP)$, a family of 
bounded multiplication operators $V_\om$ on $L^2(\RR^2)$ is defined by

$$
(V_\om \psi)(\x)\;=\;
V(-\x\cdot \om)\,\psi(\x)
\mbox{ . }
$$

\noindent The family of Hamiltonians studied here is now
$H_\omega=H_L+V_\omega$. Note that  
$H_\om$ transforms covariantly with
respect to the magnetic translations: 

\begin{equation}
\label{eq-Hamilcov}
U(\vec{\xi}\,)H_\omega U(\vec{\xi}\,)^* 
\;=\; H_{{\vec \xi}\cdot \om}
\mbox{ , }
\qquad
\vec{\xi}\in\RR^2
\mbox{ . }
\end{equation}

\noindent By functional calculus, any (continuous)
function of the Hamiltonian is also covariant.

\vspace{.2cm}

This work will mainly focus on the analysis of magnetic operators with
an infinite boundary. Such a boundary could be modeled by a confining
potential, but in this work Dirichlet boundary
conditions are chosen. Hence
let $H_{\omega,s}$ be the operator $H_\omega$ restricted to the domain
$\{ \x\in \RR^2\,|\,x_2>-s\}$ and with Dirichlet boundary conditions.
The operator $\hat H$ from the introduction was meant to be of that type.  
$H_{\om,s}$ now satisfies the
following covariance relation:

\begin{equation}
\label{eq-Hamilcovedge}
U(\vec{\xi}\,)H_{\omega,s} U(\vec{\xi}\,)^*\;=\;
H_{\vec{\xi}\cdot\om,s+\xi_2}
\mbox{ , }
\qquad
\vec{\xi}\in \RR^2
\mbox{ . }
\end{equation}

\noindent This relation can be made even more similar to
(\ref{eq-Hamilcov}) if one introduces the new 
(non-compact) hull $\hat{\Omega}=\Omega\times(\RR\cup\infty)$ 
and furnishes it with
the $\RR^2$-action
$\vec{\xi}\cdot(\om,s)=(\vec{\xi}\cdot\om,s+\xi_2)$ where $\infty+x_2=\infty$. 
As $s=\infty$ is left invariant under the shift, the point
$\omega\in\Omega$ (without boundary conditions) can be identified with the
point $(\omega,\infty)\in\hat{\Omega}$ (boundary conditions pushed to
$\infty$). In other words, equation
(\ref{eq-Hamilcovedge}) incorporates (\ref{eq-Hamilcov}).

\vspace{.2cm}

\noindent {\bf Example:}
For sake of concreteness, let us construct such a potential
explicitely. Set $\Omega= \RR^2/ \ZZ^2
\times [-\lambda,\lambda]^{\times
\ZZ^2}$ and let $\PP$ be the product measure of the Lebesgue measures with
i.i.d. measures on the $\lambda$-components. If
$\omega=({\x}_0,(\lambda_\omega(\n))_{\n\in\ZZ^2})$, then the action
is given by $\x\cdot \om=
(\x+{\x}_0,(\lambda_\omega(\n+[\x+{\x}_0]))_{\n\in\ZZ^2})$ where $[\x]$
denotes the integer parts of $\x$.  
Suppose now given two positive functions
$w,v\in L^\infty(\RR^2) $ where $w$ is periodic with unit
cell $[0,1]^2$ and 
$v$ vanishes on the boundary and outside of the unit cell. 
Then set $V(\omega)=w({\x}_0)+\lambda_\om(0)\,v({\x}_0)$. 
The associated multiplication
operator is the sum of the periodic potential and 
disordered potential of the following type:

$$
V_\omega(\x)\;=\;w(\x+{\x}_0)+
\sum_{\n\in\ZZ^2} \lambda_\om(\n)\,v(\x+{\x}_0+\n)
\mbox{ . }
$$

\noindent It is well-known that the periodic potential splits the
Landau bands, each giving Harper-like spectra, and that the disordered
potential leads to localization ({\sl e.g.} \cite{CH,GK1}, and references
therein). 
\hfill $\diamond$

\vspace{.2cm}

\section{Analysis of covariant families of integral operators}

Equation (\ref{eq-Hamilcovedge}), incorporating (\ref{eq-Hamilcov}), is
a covariance relation for the Hamiltonians  
$(H_{\hat{\omega}})_{\hat{\omega}\in\hat{\Omega}}$.
By functional calculus it leads to a covariance relation also for
functions of these operators.
As will be proven in Sections \ref{sec-kernelplane} and 
\ref{sec-halfplanekernel} below, certain functions of the Hamiltonians
$H_{\oh}$ will actually be bounded integral
operators on $L^2(\RR^2)$. Therefore this section is concerned with
the set $\Ta$ of weakly continuous
families $A=(A_{\hat{\omega}})_{\hat{\omega}\in\hat{\Omega}}$
of bounded integral operators on
$L^2(\RR^2)$ for which the covariance relation

\begin{equation}
\label{eq-cov}
U(\vec{\xi}\,)A_{\hat{\omega}} U(\vec{\xi}\,)^* 
\;=\; A_{{\vec \xi}\cdot {\hat{\omega}}}
\mbox{ , }
\qquad
\vec{\xi}\in\RR^2
\mbox{ , }
\end{equation}

\noindent holds and 

$$
\|A\|_\infty\;=\; 
\esssup_\oh \|A_\oh\|
\;<\;\infty
\mbox{ , }
$$

\noindent 
where the essential supremum is taken with respect to
the product of the probability measure $\PP$ with the
Lebesgue measure. As already mentioned above,
$\Omega\times\{\infty\}$ is an $\RR^2$-invariant compact subspace of
$\hat\Omega$ so that 
$(A_{\omega,\infty})_{\omega\in{\Omega}}$ forms also a weakly
continuous covariant family
of bounded integral operators on
$L^2(\RR^2)$. The set of these families is denoted by $\Aa$ and 
in order to simplify
notations $A_\om$ will be written for $A_{\omega,\infty}$. 

\vspace{.2cm}

Pointwise linear combinations and products of such families form new 
families, hence $\Ta$ is actually an algebra, a generalized
convolution algebra.
In fact, various crossed product
algebras are naturally associated to covariant families of integral
operators (smooth and C$^*$-algebras, as well as a von Neumann
algebra \cite{Con,Bel91}), but this point of view
will not be developed here. Let us only mention that the restriction of a
family  $(A_{\oh})_{\oh\in\hat{\Omega}}$ to values $s=\infty$
yields an algebra homomorphism $\Ta\to\Aa$ which plays an important
role in \cite{KS}. 
In the following, the standard algebraic structures
like derivations, covariant derivatives
and various traces for covariant operator families are introduced.

\vspace{.2cm}

The integral kernel of $A_\oh$ will be denoted by $\langle
\x\,|A_\oh|\y\,\rangle$. Continuity and differentiability properties 
of these kernels in $\x$ and $\y$, as
well as estimates on the decay in $\x-\y$ will be studied in the next
section. These properties transpose again to sums and products. 
Using the covariance relation and the Cauchy-Schwarz inequality,
one establishes that
\begin{equation}
\label{eq-normbound}
\|A\|_\infty 
\;\leq\; \int_\RR d\x\:
\esssup_\oh 
\left|\langle \x\,|A_\oh|\vec{0}\,\rangle
\right| 
\mbox{ . }
\end{equation}

\vspace{0.2cm}

Given $A\in\Ta$, new elements $\nabla_j A\in\Ta$, $j=1,2$, 
are defined by 

\begin{equation}
\label{eq-nabla}
(\nabla_j A)_\oh
\;=\;
\imath[X_j,A_\oh]
\mbox{ , }
\end{equation}

\noindent as long as the r.h.s. are again bounded integral operators.
By (\ref{eq-normbound}), this can be assured for through decay
properties of the kernels $\langle
\x\,|A_\oh|\y\,\rangle$ in $|\x-\y\,|$.
$\nabla_j$ is a derivation, i.e.\ it satisfies the Leibniz rule 
$\nabla_j (AB)=\nabla_j (A)B+A\nabla_j(B)$. 
Furthermore, $D_j A\in\Ta$ has integral
kernel

\begin{equation}
\label{eq-coderiv}
\langle
\x\,|D_j A_\oh|\y\,\rangle
\;=\;
\left(\imath \partial_{x_j}-\delta_{j,1}\gamma x_2\right)
\langle\x\,|A_\oh|\y\,\rangle
\mbox{ , }
\end{equation}

\noindent again provided the r.h.s. is the integral kernel of a
bounded operator. If a
covariant family $A=(A_{\omega,s})_{\omega,s}$ is actually independent
of $\omega$, then the covariance relation implies that the integral kernel
$\langle\x\,|A_{\omega,s}|\y\,\rangle$ depends only on $x_1-y_1,x_2,y_2$
and that of   $\langle\x\,|A_{\omega,\infty}|\y\,\rangle$ only on
$\x-\y$. As a consequence, for such $A$ 
\begin{equation}
\label{eq-leibnitz}
[D_1,A_{\omega,s}]
\;=\;
\imath\gamma \nabla_{2} A_{\omega,s}\mbox{ , } \qquad
[D_2,A_{\omega,\infty}] \;=\; 0 \mbox{ . }
\end{equation}

\vspace{.2cm}

Let $\chi$ be a positive compactly supported 
function on $\RR$ satisfying $\int
dx\,\chi(x)=1$. 
For $j=1,2$, let us set 
$\chi_{j}(\x)=\chi(x_j)$ and consider $\chi_{j}$ also as a
multiplication operator on $L^2(\RR^2)$.
Let $\| T\|_1$ denote the (Schatten) traceclass norm of an operator
$T$ on $L^2(\RR^2)$. Whenever 
$\|\chi_{1}A_\oh\|_1$ 
is integrable w.r.t. $\PP$, the family $A\in\Ta$ will be called 
$\TVh$-traceclass. Whenever 
$\|\chi_{1}\chi_2A_{\omega,\infty}\|_1$ is integrable w.r.t. $\PP$,
the family $A\in\Aa$ 
is called $\TV$-traceclass. For traceclass families, one can set

$$
\TVh(A)
\;=\;
\int d\PP(\omega)\;
\Tr(\chi_{1}A_{\omega,s})
\mbox{ , }
\qquad
\TV(A)
\;=\;
\int d\PP(\omega)\;
\Tr(\chi_{1}\chi_{2}A_{\omega,\infty})
\mbox{ , }
$$

\noindent where $\Tr$ is the usual trace on
$L^2(\RR^2)$.  
In order to write out more explicit formulas and see that the
definition of $\TVh$ is independent of the choice of $s$, 
recall that if $T$ is a traceclass integral operator on
$L^2(\RR^2)$ with jointly continuous integral kernel,
then $\Tr(T)=\int
d\x \,\langle \x\,|T|\x\,\rangle$ (jointly continuous means that 
$(\x,\y)\mapsto\langle \x\,|T|\y\,\rangle$ is continuous;
references herefore are given, {\sl
e.g.}, in \cite{ASS} where it is also shown that the same formula
holds if the integral kernel has a finite number of
isolated point singularities). 
Using the covariance relation
(\ref{eq-cov}) and the invariance of $\PP$, 

\begin{equation}
\label{eq-tracecalc}
\TVh(A)
\;=\;
\int d\PP(\omega)
\int ds\;
\langle \vec{0} \,|A_{\omega,s}|\vec{0}\,\rangle
\mbox{ , }
\qquad
\TV(A)
\;=\;
\int d\PP(\omega)\,
\langle \vec{0}\,|A_{\omega,\infty}|\vec{0}\,\rangle
\mbox{ . }
\end{equation}

\noindent This shows that $A$ 
is $\TVh$-traceclass (respectively $\TV$-traceclass)
if the integral kernels of $|A|$ are jointly continuous
and integrable in the $2$-direction (respectively, 
jointly continuous and uniformly bounded). 

\begin{lem}
\label{lem-traces} $\TVh$ and $\TV$ are traces on $\Ta$. 
This means in the case of $\TVh$
that for $A,B\in\Ta$ with $\TVh$-traceclass $B$: 

\vspace{.1cm}

\noindent {\rm (i)} $\TVh(AB)=\TVh(BA)$.

\vspace{.1cm}

\noindent {\rm (ii)} $\TVh(AB)\leq \|A\|_\infty\,\TVh(|B|)$.

\vspace{.1cm}

\noindent {\rm (iii)} $\TVh(|A+B|)\leq \TVh(|A|)+
\TVh(|B|)$.
\vspace{.1cm}

\noindent Similar relations hold for $\TV$.
\end{lem}

\noindent
\bew\ 
Because of the translation invariance of $\PP$,
(i) can immediately be deduced from the definition of $\TVh$. 
In order to prove (ii), one can use the polar decomposition $B=U|B|$
where the unitary $U=(U_\oh)_{\om\in\hat{\Omega}}$ 
is easily seen to satisfy the covariance
relation, just as the positive operator $|B|$. Then 

$$
\TVh (AB)\;=\;
\int d\PP(\omega)\,
\Tr(A_\oh U_\oh |B_\oh|\chi_1)
\;\leq\;
\int d\PP(\omega)\,\|A_\oh U_\oh \|\;
\Tr(|B_\oh|\chi_1)
\;\leq\;\|A\|_\infty\;\TVh (|B|)
\mbox{ . }
$$

\noindent For the proof of (iii), set $|A+B|=U(A+B)$ by polar
decomposition. Then $\TVh(|A+B|)=\TVh(UA)+\TVh(UB)$ which allows to
conclude by (ii).
\eb

\vspace{.2cm}

It follows from the covariance relation and the Birkhoff theorem that
$\TV$ is the trace per unit volume \cite{Bel91}, while $\TVh$ is the
(disorder averaged) trace per unit volume in the $1$-direction
combined with the usual trace in the $2$-direction. 
Finally let us remark that the traces are invariant w.r.t. the
derivations:

$$
\TVh(\nabla_1 A)\;=\;0
\mbox{ , }
\qquad
\TV(\nabla_j A)\;=\;0
\mbox{ , }
\quad
j=1,2
\mbox{ , }
$$

\noindent as long as $A$ has jointly
continuous integral kernel and $\nabla_1
A$ is $\TVh$-traceclass (resp. $\nabla_j
A$ is $\TV$-traceclass). Under these hypothesis, this
can directly be verified from the expressions (\ref{eq-tracecalc}).

\vspace{.2cm}

\section{Integralkernels associated to the planar Hamiltonian}
\label{sec-kernelplane}

In this section, Hamiltonians without boundary conditions are
considered. Hence $s=\infty$ and $\omega$ stands for $(\omega,\infty)$.
Following Davies \cite[Sec. 3.4]{Dav},
the functional calculus of the Hamiltonian
$H_\omega=H_L+V_\omega$ will be done via the complex heat kernel:

\begin{equation}
\label{eq-fctcalc}
F(H_\omega)
\;=\;
\int^\infty_{-\infty} dt\,\tilde{F}(t)\,e^{-H_\omega(1+\imath t)}
\mbox{ , }
\end{equation}

\noindent where

$$
\tilde{F}(t)\;=\;2\pi\,
\int^\infty_{-\infty} dE\,e^{\imath Et}\,e^E\,F(E)
\mbox{ . }
$$

\noindent For compactly supported differentiable  
functions $F\in C^{k}_c(\RR)$, one has  
the standard Fourier estimates  $|\tilde{F}(t)|\leq
c_k(1+|t|^{k-1})^{-1}$. Such an estimate may also hold for functions
with infinite support, but we do not intend here to give the most
general formulation.

\begin{prop}
\label{prop-Hamilfunction}
Let $V_\om\in L^\infty(\RR^2)$
and $F\in C^{k}_c(\RR)$ with $k>2$. Then
$F(H_\omega)$ is an integral operator the integral
kernel of which satisfies uniformly in $\omega$ and
for any $\delta>0$

$$
|\langle\x|F(H_\omega)|\y\,\rangle|
\;\leq\;
\frac{c_\delta}{1+|\x-\y\,|^{k-2-\delta}}
\mbox{ . }
$$

\end{prop}

\noindent
\bew\ 
As the following estimates are pointwise in
$\omega$, the index will be suppressed. Let us begin with
the integral kernel of $e^{-H_Lz}$
explicitly using Mehler's formula for the (shifted) harmonic oscillator
$h(k)=\frac{1}{2}(-\partial_{x_2}^2+
\gamma^2(X_2+\frac{k}{\gamma})^2)$ 
in the 2-direction ($\Re e(z)>0$): 
\begin{eqnarray}
\langle\x|e^{-H_Lz}|\y\,\rangle
& = &
\int \frac{dk}{2\pi}\;
e^{\imath k(x_1-y_1)}
\langle x_2|e^{-h(k)z}|
y_2\rangle
\nonumber
\\
&  &
\label{eq-Landaukernel}
\\
& = &
\frac{\gamma}{4\pi}\;\frac{1}{\mbox{sinh}(\frac{\gamma}{2}z)}
\;e^{-\frac{\gamma}{4}\coth(\frac{\gamma}{2}z)|\x-\y|^2}
\;e^{-\imath\frac{\gamma}{2}(x_1-y_1)(x_2-y_2)}
\;\Phi(\x-\y,\y\,)
\mbox{ . }
\nonumber
\end{eqnarray}

\noindent 
In order to obtain upper bounds for the integral kernel
let us use the following elementary inequalities ($\Re e(z)>0$):

\begin{equation}
\frac{1}{|\sinh(z)|}\;\leq\;\frac{1}{\Re e(z)}
\mbox{ , }
\qquad
\Re e(\coth(z))\;\geq\;\Re e(z^{-1})
\mbox{ . }
\label{eq-basicineq}
\end{equation}

\noindent They lead directly to the following estimate:

\begin{equation}
\left|
\langle\x|e^{-H_Lz}|\y\,\rangle\right|
\;\leq\;
\frac{1}{2\pi}\;\frac{1}{\Re e(z)}
\;
e^{-\frac{1}{2}|\x-\y\,|^2\,\Re e(z^{-1})}
\mbox{ . }
\label{eq-Landauest}
\end{equation}

\noindent Now set $V_+=\|V\|_\infty$ and $\tilde{V}=V-V_+$ so that
$\tilde{H}=H_L+\tilde{V}$ has a negative potential $\tilde V$.
Furthermore Duhamel's formula reads

\begin{equation}
\langle\x|e^{-\tilde{H}z}|\y\,\rangle
\;=\;
\langle\x|e^{-H_Lz}|\y\,\rangle
-z\int^1_0 dq\;\int_{\RR^2} d\vec{r}\;
\langle\x\,|e^{-(1-q)H_Lz}|\vec{r}\,\rangle\,\tilde{V}(\vec{r}\,)
\,\langle\vec{r}\,|e^{-q\tilde{H}z}|\y\,\rangle
\mbox{ . }
\label{eq-Duham}
\end{equation}

\noindent Using this iteratively, one obtains the Dyson series for
$z=t>0$ which
is estimated term by term using (\ref{eq-Landauest})

\begin{equation}\label{J-eq3}
\left|\langle\x|e^{-Ht}|\y\,\rangle\right|
\;\leq\;
e^{-tV_+}\,\sum_{n\geq 0} \frac{V_+^n}{(2\pi)^n} 
\left(\prod_{l=1}^n\int_{\RR^2} d\vec{r}_l\int^{q_{l+1}}_0
\!\!\!dq_l\right)
\left(\,\prod_{l=0}^{n-1}
\frac{e^{-\frac{|\vec{r}_{l-1}-\vec{r}_l\,|^2}{2(q_l-q_{l-1})t}}}{q_l-q_{l-1}}
\,\right)
\frac{e^{-\frac{|\vec{r}_{n}-\y\,|^2}{2q_nt}}}{q_nt}
\mbox{ ,}
\end{equation}

\noindent where $q_{n+1}=1$ and $\vec{r}_0=\x$ 
in each term. A short calculation using 
rotation invariance shows, for $\Re e(a)>0$ and  $\Re e(b)>0$,

\begin{equation}
\int_{\RR^2} d\vec{r}\;e^{-\frac{|\x-\vec{r}\,|^2}{a}}\,
e^{-\frac{|\vec{r}-\y\,|^2}{b}}
\;=\;
\pi\,\frac{ab}{a+b}\,e^{-\frac{|\x-\y\,|^2}{a+b}}
\mbox{ . }
\label{eq-integral}
\end{equation}

\noindent Applying this $n$ times in the $n$th order term of the Dyson series
shows

\begin{equation}
\left|\langle\x|e^{-Ht}|\y\,\rangle\right|
\;\leq\;
\frac{1}{2\pi\,t}\;
e^{-\frac{1}{2t}|\x-\y\,|^2}
\mbox{ . }
\label{eq-semigroupest}
\end{equation}

\noindent Now the arguments of Lemma 3.4.6 and Theorem 3.4.8 
of \cite{Dav} imply that, for $\Re e(z)>0$,

\begin{equation}
\label{eq-complexest}
|\langle \x|e^{-zH}|\y\,\rangle|
\;\leq\;
\frac{c}{|\Re e(z)|}\,e^{-\frac{1}{4}|\x-\y\,|^2\,\Re e(z^{-1})}
\mbox{ . }
\end{equation}

\noindent As in Theorem
3.4.9 of \cite{Dav}, one therefore has

$$
|\langle\x|F(H_\omega)|\y\,\rangle|
\;\leq\;
\int dt \;\frac{c_k}{1+|t|^{k-1}}\,
\exp\left(-\frac{1}{4}\,\frac{|\x-\y\,|^2}{1+t^2}\right)
\mbox{ , }
$$

\noindent so that the inequality $e^{-r}\leq c_\beta/(1+r)^\beta$ for
$r,\beta>0$ leads to

$$
|\langle\x|F(H_\omega)|\y\,\rangle|
\;\leq\;
\int dt \;\frac{c_k\,c_\beta}{1+|t|^{k-1}}\,(1+t^2)^\beta\,
\frac{1}{(1+\frac{1}{4}|\x-\y\,|^2)^\beta}
\mbox{ . }
$$

\noindent Hence the $t$-integral is bounded as long as $2\beta<k-2$
which concludes the proof. 
\eb

\vspace{.2cm}

As an aside be mentioned that
there are various other ways to get estimates on the integral kernel
of the semigroup $e^{-t H}$. One is a Combes-Thomas-like argument
which will be used in Section \ref{sec-halfplanekernel}. Another
is to simply apply the
diamagnetic inequality 
\cite[Theo.~1.13]{CFKS}, which reads $|e^{-tH}\phi(\x)|\leq 
e^{t\Delta/2}|\phi|(\x)$ for positive $V$ and 
any $\phi \in L^2(\RR^2)$ and $t>0$ where
$\Delta=\partial_1^2+\partial_2^2$ is the two-dimensional Laplacian.
As is moreover known (consult {\sl e.g.} \cite{BHL})
that the integral kernels $\langle
\x|e^{-tH}|\y\,\rangle$ are jointly continuous,
one also deduces the pointwise estimate
(\ref{eq-semigroupest}) because the r.h.s. of (\ref{eq-semigroupest})
is precisely the integral kernel of $e^{t\Delta/2}$.
Here the above Dyson series argument was used because the
same technique will be used to derive estimates on the
covariant derivatives of the integral kernels. 

\begin{prop}
\label{prop-Hamilderive}
Let $V_\omega\in L^\infty(\RR^2)$ 
and $F\in C^{k}_c(\RR)$ with $k>6$. Then 
$D_j F(H_\omega)$
is an integral operator satisfying
for any $\delta>0$

$$
|\langle\x|D_j F(H_\omega)|\y\rangle|
\;\leq\;
\frac{c_\delta}{1+|\x-\y\,|^{k-6-\delta}}
\mbox{ . }
$$

\noindent
Suppose that $\partial_j V_\omega\in L^\infty(\RR^2)$ and
$F\in C^{k}_c(\RR)$ with $k>10$. Then 
$D_j D_i F(H_\omega)$
is an integral operator satisfying
for any $\delta>0$

$$
|\langle\x|D_jD_i F(H_\omega)|\y\rangle|
\;\leq\;
\frac{c_\delta}{1+|\x-\y\,|^{k-10-\delta}}
\mbox{ . }
$$

\end{prop}

\vspace{.2cm}

\noindent
\bew\ Again the index $\omega$ will be suppressed. 
One has

\begin{equation}
D_je^{-zH}
\;=\;
D_je^{-zH_L}
-z\,
\int^1_0 dq\;
D_j e^{-z(1-q)H_L}
\;
V\,e^{-zqH}
\mbox{ . } 
\label{eq-help}
\end{equation}

\noindent Hence estimates on the covariant derivatives of the Landau
Hamiltonian will be needed. Using
$|\coth(z)|\leq\frac{e^{-\Re e(z)}}{\Re e(z)}+1$ and the inequalities
(\ref{eq-basicineq}) (from now on $c$ denotes varying constants and
$\Re e(z)\geq 0$)

\begin{eqnarray}
\left|(\imath\partial_{x_j}-\gamma\delta_{j,1}x_2)\,
\langle\x|e^{-z H_L}|\y\,\rangle\right|
& \leq  & 
\frac{c}{\Re e(z)}\;|\x-\y\,|\;\left(1+
\frac{1}{\Re e(z)}\right)
\;
e^{-\frac{1}{4}|\x-\y\,|^2\,\Re e(z^{-1})}
\nonumber
\\
& &
\label{eq-Landauderiv}
\\
& \leq & 
\frac{c}{\Re e(z)^2\sqrt{\Re e(z^{-1})}}\;\left(\Re e(z)+
1\right)
\;
e^{-\frac{1}{8}|\x-\y\,|^2\,\Re e(z^{-1})}
\mbox{ , }
\nonumber
\end{eqnarray}

\noindent where in the second step 
$ae^{-2a^2}\leq e^{-a^2}$ for $a>0$ was used.
Let now 
$I_1$ denote the integral kernel of the second contribution in
(\ref{eq-help}).  Using
(\ref{eq-complexest}) and then again (\ref{eq-integral}), one gets 
\begin{eqnarray}
|I_1| \!\! & \leq  & \!\!c\,
\frac{\Re e(z)+1
}{\Re e(z)^2 \sqrt{\Re e(z^{-1})}}\,|z|\,
\int^1_0 \!dq
\int\! d\vec{r}\,
\frac{1}{(1-q)^{\frac{3}{2}}}\,
e^{-\frac{1}{8(1-q)}|\x-\vec{r}\,|^2\,\Re e(z^{-1})}
\,\frac{1}{q\Re e(z)}\,
e^{-\frac{1}{8q}|\vec{r}-\y\,|^2\,\Re e(z^{-1})}
\nonumber
\\
& &
\label{eq-firstderiv}
\\
& \leq & \!\!c\;
\frac{\Re e(z)+1
}{\Re e(z)^3 \sqrt{\Re e(z^{-1})}}
\;\frac{|z|}{\Re e(z^{-1})}\;
e^{-\frac{1}{8}|\x-\y\,|^2\,\Re e(z^{-1})}
\nonumber
\mbox{ . }
\end{eqnarray}

\noindent Hence, 
\begin{equation}
\left|
\langle\x|D_je^{-zH}|\y\,\rangle\right|
\;\leq\;
c\,
\frac{\Re e(z)+1
}{\Re e(z)^2 \sqrt{\Re e(z^{-1})}}
\,\left(1+\frac{|z|}{\Re e(z)\Re e(z^{-1})}\right)\,
e^{-\frac{1}{8}|\x-\y\,|^2\,\Re e(z^{-1})}
\mbox{  }
\label{eq-compsemi}
\end{equation}

\noindent 
yielding
\begin{equation}
\left|
\langle\x|D_je^{-(1+\imath t)H}|\y\,\rangle\right|
\;\leq\;
c\,(1+t^2)^2 
e^{-\frac{|\x-\y\,|^2}{8(1+t^2)}}
\mbox{ . }
\end{equation}

\noindent 
This implies just as in Proposition 
\ref{prop-Hamilfunction}
that $D_j F(H)$ satisfies the stated bound. 

\vspace{.2cm}

To prove the second statement, let us 
use (\ref{eq-leibnitz}) which implies

\begin{eqnarray}
D_j D_i e^{-zH} & =  & 
D_j D_i e^{-zH_L} 
+\delta_{i,1}\;\imath\gamma z\,
\int^1_0 dq 
\,D_j\left(\nabla_{2} e^{-(1-q)zH_L}\right)
Ve^{-qzH}
\nonumber
\\
& &
\nonumber
\\
&  & +\,\imath z\,
\int^1_0 dq\; D_je^{-(1-q)zH_L}\partial_i V 
e^{-qzH}
\,+\,z
\int^1_0 dq \;D_je^{-(1-q)zH_L}V D_i
e^{-qzH}
\nonumber
\mbox{ . }
\end{eqnarray}

\noindent As in (\ref{eq-Landauderiv}) one shows for the first term

$$
\left|
\langle\x|D_j D_i e^{-zH_L}|\y\,\rangle\right|
\; \leq  \; 
c\;\frac{\left(\Re e(z)+1\right)^2}{\Re e(z)^{3}}
\;
\frac{1}{\Re e(z^{-1})}
\;
e^{-\frac{|\x-\y\,|^2}{16}\,\Re e(z^{-1})}\;
\mbox{ . }
$$

\noindent
For the second term, let us commute $D_j$ and $\nabla_2$. The integral
kernel of the contribution $\int^1_0 dq 
[D_j,\nabla_2] e^{-(1-q)zH_L}Ve^{-qzH}$ satisfies a bound as 
(\ref{eq-firstderiv}). Let $I_2$ be the integral kernel of $\int^1_0 dq 
\nabla_{2}D_j e^{-(1-q)zH_L}Ve^{-qzH}$. Using
$$ \left|
\langle\x|\nabla_2 D_j e^{-zH_L}|\y\,\rangle\right|
\, \leq  \,
c\;\frac{\left(\Re e(z)+1\right)^2}{\Re e(z)^{2}}
\,|\x-\y\,|^2
e^{-\frac{|\x-\y\,|^2}{4}\,\Re e(z^{-1})}
\, \leq  \, 
\frac{c\left(\Re e(z)+1\right)^2}{\Re e(z)^{2}\Re e(z^{-1})}\;
e^{-\frac{|\x-\y\,|^2}{16}\,\Re e(z^{-1})}
\mbox{ , }
$$
and performing a similar calculation as in
(\ref{eq-firstderiv})
one finds that it can be bounded by
$$ |I_2| \; \leq  \; 
c\;\frac{|z|\left(\Re e(z)+1\right)^2}{\Re e(z)^{3}\Re e(z^{-1})^2}\;
e^{-\frac{|\x-\y\,|^2}{16}\,\Re e(z^{-1})}
\mbox{  }
$$
The integral kernel of the third contribution can be bounded as above,
hence let us focus on the integral kernel $I_4$ of the forth contribution.
Using (\ref{eq-Landauderiv})
and (\ref{eq-compsemi}), one gets by a similar calculation as in
(\ref{eq-firstderiv})

$$
|I_4| \; \leq  \;c\;
\frac{(\Re e(z)+1)^2}{\Re e(z)^4 \Re e(z^{-1})}
\left(1+\frac{|z|}{\Re (z)\Re e(z^{-1})}\right)\;\frac{|z|}{
\Re e(z^{-1})}\;
e^{-\frac{|\x-\y\,|^2}{16}\,\Re e(z^{-1})}
\mbox{ . }
$$

\noindent
Finally, the number $k$ determining the decay of the
integral kernel of $D_jD_iF(H_\omega)$ depends on the leading power in
$t$ of  $\left|\langle\x|D_j D_i e^{-(1+\imath t)H}|\y\,\rangle\right|$.
Comparing the above contributions one sees that this power is
determined by $I_4$, and, setting $z=1+\imath t$, one has

$$
|I_4|
\;\leq\;
c\;(1+t^2)^{4}\;
e^{-\frac{|\x-\y\,|^2}{16(1+t^2)}}
\mbox{ . }
$$

\noindent As in  Proposition~\ref{prop-Hamilfunction}, 
the statement of the proposition follows.
\eb

\vspace{.2cm}

Let us remark that Proposition \ref{prop-Hamilderive}
implies in particular that the integral kernel of
$F(H_\omega)$ is twice differentiable.
In dimension 2, the same argument goes through for 
$D_1^2D_2^2\,F(H)$, but not for $D_j^3\,F(H)$. In higher dimension, more
regularity can be obtained.

\section{Integralkernels of operators on the half-plane}
\label{sec-halfplanekernel}

The aim of this section is to show that 
Proposition~\ref{prop-Hamilfunction} and the part of 
Proposition~\ref{prop-Hamilderive} concerning covariant derivatives in
the $1$-direction remain essentially valid 
for the operators $H_{\omega,s}$ on the
half-plane. This is done by proving estimates like
(\ref{eq-Landauest}) and (\ref{eq-Landauderiv}) 
for the kernel of the semigroup generated by
the Landau operator $\hat{H}_L$ (and its covariant derivative)
with Dirichlet boundary conditions at
$s=0$. Covariance then implies that these estimates also hold for arbitrary
$s<\infty$ and the perturbative arguments based on the Dyson series
expansion can be directly transposed to obtain a power law decay of the
integral kernels of functions of the Hamiltonian on the half-plane.

\vspace{.2cm}

\begin{prop}
\label{prop-halfplaneLandau}
For $\Re e(z)>0$ and $n=0,1,2$. Then

$$
\left|\;
\langle\x\,|D_1^n\,e^{-z\hat{H}_L}|\y\,\rangle
\;\right|
\;\leq\;
c\,\frac{1+|z|^{n+1}}{\Re e(z)^{\frac{n}{2}+1}}
\exp\left(-\frac{|\x-\y|^2}{10}\,\Re e(z^{-1})\right)
\mbox{ . }
$$

\noindent 
Furthermore, $\langle\x\,|D_1^n e^{-z\hat{H}_L}|\y\,\rangle$ is 
continuous in $\x,\y$ for $n=0,1$.
\end{prop}

This in particular implies that the integral kernel
of $e^{-t H_{\omega,s}}$ is continuous so that the diamagnetic inequality
implies

\begin{equation}
\label{eq-diamaghalf}
\left|\langle \x|e^{-t H_{\omega,s}}|\y\,\rangle\right|
\;\leq\;
\langle \x|e^{t\Delta_s}|\y\,\rangle
\;=\;
\frac{1}{4\pi t}\,e^{-|\x-\y\,|^2/t}
\left(1-e^{-2(x_2+s)(y_2+s)/t}\right)
\;\chi(x_2\geq -s)\,\chi(y_2\geq -s)
\mbox{ . }
\end{equation}

\noindent 
This also shows
how the integral kernels of functions of $H_{\omega,s}$
vanish near $x_2=-s$ or $y_2=-s$. 

\vspace{.2cm}

For the proof of Proposition \ref{prop-halfplaneLandau},
the semigroup of $\hat{H}_L$ is calculated via
Fourier transform just as in (\ref{eq-Landaukernel}):

\begin{equation}
\label{eq-Landauhalf}
\langle\x\,|e^{-t\hat{H}_L}|\y\,\rangle
\; = \;
\int \frac{dk}{2\pi}\;
e^{\imath k(x_1-y_1)}\;
\langle x_2|e^{-t\hat{h}(k)}|
y_2\rangle
\mbox{ , }
\end{equation}

\noindent where $\hat{h}(k)=\frac{1}{2}(-\partial^2+
\gamma^2(X+\frac{k}{\gamma})^2)$ with Dirichlet boundary conditions at
the origin. As we did not
succeed in calculating this kernel explicitly, recourse to more
abstract analytical arguments is necessary.
For a complex dilation argument on the heat kernel, the
following will be needed:

\begin{lem}
\label{lem-analytic}
$k+\imath\kappa
\in\CC\mapsto e^{-t\hat{h}(k+\imath\kappa)}$ 
is entire for all $t> 0$ and the
integral kernel satisfies

\begin{equation}
\label{eq-mehlercomplex}
\left|
\langle x|e^{-t\hat{h}(k+\imath\kappa)}|
y\rangle
\right|
\;\leq\;
e^{\frac{1}{2}\,t\kappa^2}
\;\langle x|e^{-t h(k)}|y\rangle
\mbox{ . }
\end{equation}

\end{lem}

\noindent
\bew\ 
First let us show that $X$ is
relatively bounded w.r.t. $\hat{h}(0)$ with relative bound
$0$. Therefore let $|n\rangle$ denote the Hermite eigenfunctions 
of $h(0)$ and recall $X|n\rangle=(2\gamma)^{-1/2}
(\sqrt{n+1}\,|n+1\rangle+\sqrt{n}\,|n-1\rangle)$. The odd Hermite functions
$|2l+1\rangle$ form an eigenbasis of $\hat{h}(0)$ which is
complete in $L^2(\RR_+)$. Now let $\psi=\sum_{l\geq 0}
a_l\,|2l+1\rangle$ so that $\|\hat{h}(k)\psi\|^2=
\sum_{l\geq 0}|a_l|^2(2l+\frac{3}{2})^2$.
As $\|X\psi\|^2\leq c\sum_{l\geq 0}|a_l|^2
(2l+2)$, the relative bound estimates follow immediately.
In conclusion, $\hat{h}(k+\imath\kappa)=\hat{h}(0)+\gamma X
(k+\imath\kappa)+\frac{1}{2}(k+\imath\kappa)^2$ is
automatically closed and \cite[Thm.~IX.2.6]{Kat} implies the desired 
analyticity property.

\vspace{.2cm}

In order to prove the estimate, let us cite a norm-convergent
version of the Trotter product formula from \cite{CZ}:
Given two $m$-sectorial operators
$A,B$ on a given Hilbert space ${\cal H}$ 
satisfying that $(A+1)^{-1}$ is compact
and $D(A)\cap D(B)$ is dense in ${\cal H}$,

$$
e^{-t(A\dot{+} B)}
\;=\;
\lim_{n\to\infty}\left(e^{-\frac{t}{n}A} e^{-\frac{t}{n}B}\right)^n 
\mbox{ , }
$$

\noindent where the convergence is 
in the norm topology and $A\dot{+} B$ is the form sum.
This will be applied for $A=\hat h(k)-\alpha W(k)$
and $B =\alpha W(k)  - \imath\gamma \kappa(X+\frac{k}{\gamma})$
where $W(k)=\frac{\gamma^2}{2}(X+\frac{k}{\gamma})^2+1$. 
Indeed, $A$ is a strictly
positive selfadjoint operator 
with compact resolvent
as long as $\alpha>0$ is small enough and $B$ is $m$-sectorial. 
As $A\dot{+} B-\frac{1}{2}\kappa^2$ and $\hat{h}(k+\imath \kappa)$
coincide on the domain of the latter and the semigroups are bounded,
one deduces

$$ 
e^{-t\hat{h}(k+\imath \kappa)} 
\;=\; 
e^{\frac{1}{2}\,t\kappa^2}\;\lim_{n\to\infty}
\left(e^{-\frac{t}{n}A}e^{-\frac{t}{n}B}\right)^n 
\mbox{ . }
$$

\noindent Setting $x=r_0$ and $y=r_n$, one can
therefore bound as follows:

\begin{eqnarray*}
\left| 
\langle x |\left(e^{-\frac{t}{n}A}e^{-\frac{t}{n}B}\right)^n|
y\rangle \right| 
& \leq &
\int_{\RR_+^{n-1}} dr_1\cdots dr_{n-1} 
\prod_{j=1}^n\;
\langle r_{j-1}| e^{-\frac{t}{n}(\hat h(k)-\alpha W(k))}
|r_j\rangle
\,e^{-\frac{t}{n}\alpha W(k)(r_j)}
\\ 
& = &
\langle x |\left( e^{-\frac{t}{n}(\hat h(k)-\alpha W(k))}
e^{-\frac{t}{n}\alpha W(k)}\right)^n|y\rangle 
\\ 
& = &
\langle x | e^{-t\hat h(k)}| y\rangle
\mbox{ , }
\end{eqnarray*}

\noindent where the last equality follows from recomposing with the
Trotter formula. To conclude, one just notes that the integral kernel
of $e^{-t\hat h(k)}$ is bounded by that of $e^{-th(k)}$
(this
follows easily, e.g., from the Feynman-Kac path-integral in which
Dirichlet boundary conditions are incorporated by characteristic
functions).
\eb

\vspace{.2cm}

\begin{lem}
\label{lem-semigroupreal}
For $t> 0$ and $n=0,1,2$,

$$
\left|\;
D_1^n\,
\langle\x\,|e^{-t\hat{H}_L}|\y\,\rangle
\;\right|
\;\leq\;
c\,
\frac{ (1+t)^{n+\frac{1}{2}} }{ t^{\frac{n}{2}+1} }
\exp\left(-\frac{|\x-\y|^2}{2^{n+1}t}\right)
\mbox{ . }
$$
\end{lem}

\noindent
\bew\ 
Applying $D_1^n$ to equation (\ref{eq-Landauhalf}) and 
multiplying it with $e^{\kappa(x_1-y_1)}$, $\kappa\in\RR$ leads to

$$
e^{\kappa(x_1-y_1)}\;
D_1^n\,
\langle\x\,|e^{-t\hat{H}_L}|\y\,\rangle
\; = \;
\int_\RR \frac{dk}{2\pi}\;
(-k-\gamma x_2)^n\;
e^{\imath (k-\imath \kappa)(x_1-y_1)}
\langle x_2|e^{-t\hat{h}(k)}|
y_2\rangle
\mbox{ . }
$$

\noindent Let us change variables $k-\imath\kappa\mapsto k$,
then use analyticity (Lemma \ref{lem-analytic}) and decay properties
on the boundaries of a Cauchy contour in order to obtain:

\begin{equation}
\label{eq-firstest}
e^{\kappa(x_1-y_1)}\;
D_1^n\,
\langle\x\,|e^{-t\hat{H}_L}|\y\,\rangle
\; = \;
\int_\RR \frac{dk}{2\pi}\;
(-k-\imath\kappa-\gamma x_2)^n\;
e^{\imath k(x_1-y_1)}
\langle x_2|e^{-t\hat{h}(k+\imath\kappa)}|
y_2\rangle
\mbox{ . }
\end{equation}

\noindent Now the estimate (\ref{eq-mehlercomplex}) will be used,
along with the fact $\langle x |e^{-th(k)}|y\rangle
=\langle x+\frac{k}{\gamma} |e^{-th(0)}|y+\frac{k}{\gamma}\rangle$ and
the following estimate for the Mehler kernel:

\begin{eqnarray}
\langle x |e^{-th(0)}|y\rangle
& = &
\sqrt{\frac{\gamma}{2\pi\sinh(\gamma t)}}
\;\exp
\left(
-\frac{\gamma}{4}\coth\left(\frac{\gamma t}{2}\right)|x-y|^2
-\frac{\gamma}{4}\tanh\left(\frac{\gamma t}{2}\right)|x+y|^2
\right)
\nonumber
\\
& & 
\nonumber
\\
& \leq &
\sqrt{\frac{1}{2\pi t}}\;
\;\exp
\left(
-\frac{1}{2t}|x-y|^2
-\frac{\gamma}{4}\tanh\left(\frac{\gamma t}{2}\right)|x+y|^2
\right)
\nonumber
\mbox{ . }
\end{eqnarray}

\noindent Replacing this and substituting
$k$ for $k+\frac{\gamma}{2}(x_2+y_2)$, one obtains

\begin{equation}
\left|\;e^{\kappa(x_1-y_1)}\;
D_1^n\,
\langle\x\,|e^{-t\hat{H}_L}|\y\,\rangle\;
\right|
\leq
\int \frac{dk}{2\pi}
\,\left|k+\frac{\gamma}{2}(x_2-y_2)+\imath\kappa\right|^n
\frac{e^{\frac{1}{2}t\,\kappa^2}}{\sqrt{2\pi t}}\;
\;
e^{-\frac{|x_2-y_2|^2}{2t}
-\frac{1}{\gamma}\tanh(\frac{\gamma t}{2})k^2}
\nonumber\mbox{ . }
\end{equation}

\noindent Now let us choose $\kappa=\frac{(x_1-y_1)}{t}$ and integrate over
$k$. Then

$$ 
|\langle \x | e^{-t\hat H_L} |\y\,\rangle|  
\;\leq\;  
\left(\frac{\pi\gamma}{t\tanh \left( \frac{\gamma t}{2}\right)}
\right)^{\frac{1}{2}}
e^{-\frac{|\x-\y|^2}{2t}} 
\;\leq\; 
c\;\frac{(1+t)^{\frac{1}{2}}}{t}
\;e^{-\frac{|\x-\y|^2}{2t}}
\mbox{ , }
$$

\noindent because $\coth (t) <\frac{1+t}{t}$.
Using
$\int dk\, |k-b|\,e^{-ak^2} \leq a^{-1}+\sqrt{\frac{\pi}{a}}\,|b|$ 
for $a>0$ it follows that 
 
\begin{eqnarray*} 
|D_1 \langle \x | e^{-t\hat H_L} |\y\,\rangle|
&\leq &
\left(\frac{\gamma^{\frac{1}{2}}}{(\tanh \left(
\frac{\gamma t}{2}\right))^{\frac{1}{2}} }
+ \frac{\sqrt{\pi}|x_1-y_1|}{t} +\gamma \frac{\sqrt{\pi}|x_2-y_2|}{2}\right)
\frac{\gamma^\frac{1}{2}}{
\left(t\tanh \left(\frac{\gamma t}{2}\right)\right)^{\frac{1}{2}} }
e^{-\frac{|\x-\y|^2}{2t}} \\& \leq & 
c\;\frac{(1+t)^{\frac{3}{2}}}{t^{\frac{3}{2}}}
\;e^{-\frac{|\x-\y|^2}{4t}} 
\end{eqnarray*}

\noindent
where for the second bound 
$xe^{-2x^2}\leq e^{-x^2}$ was used. 
The last estimate ($n=2$) is obtained similarly upon
using $\int dk (k-b)^2e^{-ak^2} = (b^2+2a^{-1})\sqrt{\frac{\pi}{a}}$.
\eb

\vspace{.2cm}

\noindent
{\bf Proof} of Proposition \ref{prop-halfplaneLandau}
(This argument follows closely \cite[Thm.~3.4.8]{Dav} and is hence
kept sketchy).
Let us set 
$K(z,\x,\y\,)=\langle\x\,|D^n\,e^{-z\hat{H}_L}|\y\,\rangle$. 
If $z=t+\imath s$, one has 

$$
|K(z,\x,\y\,)|
\;\leq\;
\left\|D^n\,e^{-z\hat{H}_L}\right\|_{\infty,1}
\,\leq\;
\left\|D^n\,e^{-\frac{t}{2}\hat{H}_L}\right\|_{\infty,2}
\left\|e^{-\frac{t}{2}\hat{H}_L}\right\|_{2,1}
\,=\;
\left\|D^n\,e^{-\frac{t}{2}\hat{H}_L}\right\|_{\infty,2}
\left\|e^{-\frac{t}{2}\hat{H}_L}\right\|_{\infty,2}
\mbox{.}
$$

\noindent Since
$\left\|A\right\|_{\infty,2}^2\leq \sup_{\x} 
\int_{\RR\times\RR_+}d\y \,|\langle \x |A| \y\,\rangle|^2$, 
Lemma \ref{lem-semigroupreal} implies

$$
\left\|D^n\,e^{-\frac{t}{2}\hat{H}_L}\right\|_{\infty,2}\leq
c\; \frac{(1+t)^{n+\frac{1}{2}}}{t^\frac{n+1}{2}}
$$ 

\noindent so that

$$
|K(z,\x,\y\,)|
\;\leq\;
c \,f(t)\mbox{ , }
\qquad
f(z)\;=\;\frac{1+z^{n+1}}{z^{\frac{n}{2}+1}}
\mbox{ . }
$$

Now for $0\leq\gamma<\frac{\pi}{2}$,  
let $D=\{z\,|\,0\leq \mbox{arg}(z)\leq \gamma\mbox{ , }|z|\geq 1\}$
and set

$$
g(z)\;=\;
\frac{1}{f(z^{-1})}\,K(z^{-1},\x,\y\,)\;
\exp\left(
\frac{1}{8}\,|\x-\y\,|^2 e^{\imath(\frac{\pi}{2}-\gamma)}\,
\frac{z}{\sin(\gamma)}
\right)
\mbox{ . }
$$

\noindent The hypothesis of the Phragmen-Lindel\"of
Theorem can be verified, showing that $|g(z)|\leq
c\,\cos(\gamma)^{-\frac{n+1}{2}}$ for $z\in D$. Applying this also to
$\overline{z}$ and choosing
$\gamma=\frac{\pi}{2}(1-\epsilon)+\epsilon|\arg(z)|$ for some
$\epsilon<1$ allows to
conclude the first statement of the Proposition.

\vspace{.2cm}

In order to prove continuity in $\x,\y$ of 
$\langle\x\,|D_1^n e^{-z\hat{H}_L}|\y\,\rangle$ for $n=0,1$,  
one may follow the same strategy as above to obtain a bound on 
$\langle\x\,|D_1^n\,D_2 e^{-z\hat{H}_L}|\y\,\rangle$. This involves
calculating
the derivative of the half-sided Mehler kernel with Duhamel's formula,

$$
\partial_x\,
\langle x|e^{-t\hat{h}(k)}|y\rangle
\;=\;
\int^1_0 dq
\int^\infty_0 dr\;
\partial_x\,
\langle x|e^{-(1-q)t\hat{h}(0)}|r\rangle
\;\gamma kr\;
\langle r|e^{-qt\hat{h}(k)}|y\rangle
\;
e^{-\frac{1}{2}(1-q)tk^2}
\mbox{ , }
$$

\noindent which can be done exactly as the kernel of $e^{-t\hat{h}(0)}$ is
known explicitly by the reflection principle.
One then replaces in (\ref{eq-firstest}), carries out the $k$-integral and
uses

$$
\int_\RR dr\;
r^pe^{-\frac{(r-d)^2}{(1-q)t}}e^{-\frac{r^2}{qt}}
\;\leq\;
c\,e^{-\frac{d^2}{t}}\,
(1-q)^{\frac{1}{2}}q^{\frac{1}{2}}
t^{\frac{1}{2}}(1+t^{\frac{p}{2}})
\mbox{ , }
$$

\noindent to bound the $r$-integral. Application of the inequality
$xe^{-2x^2}\leq e^{-x^2}$ allows to obtain an expression which is integrable
in $q$ at $0$ and $1$. This yields an estimate similar to but  more
cumbersome than the ones in Lemma~\ref{lem-semigroupreal}. 
Since only the continuity result is needed here, further details
are left out.
\eb

\vspace{.2cm}

\section{Comparing integral kernels}
\label{sec-lifting}

For a given function $F$, one can compare the integral kernels of
$F(H_{\omega,s})$ and $F(H_{\omega,\infty})$ and estimate the difference
in particular for arguments which are far from the boundary at
$x_2=-s$. 
Therefore let us construct the semigroup of $H_{\omega,s}$ by means of 
the reflection principle. The reflection 
$S_s:L^2(\RR^2)\to L^2(\RR^2)$ at the line $x_2=-s$ is defined by
$(S_s\psi)(x_1,x_2)=\psi(x_1,-x_2-2s)$. Let $\Pi_2^s$ be the indicator
function on the half-plane $x_2\geq -s$. Note that $S_sH_LS_s$ is 
the Landau operator with reversed magnetic field. Now set

$$
\tilde{H}_{\omega,s}\;=\;\Pi_2^s H_\omega+(1-\Pi_2^s)
S_sH_\omega S_s
$$ 

\noindent
with core $C^\infty_{c,s}(\RR^2)$, given by the functions 
in $\psi\in C^\infty_c(\RR^2)$ satisfying the antisymmetry relation
$S_s\psi=-\psi$. These functions vanish on
the boundary $x_2=-s$.
By construction, $S_s\tilde{H}_{\omega,s}S_s
=\tilde{H}_{\omega,s}$ and 
therefore $C^\infty_{c,s}(\RR^2)$ is left invariant. 
Moreover, if $\psi$ is a smooth compactly supported function in the
domain of $\tilde{H}_{\omega,s}$ then
$\tilde{H}_{\omega,s}\psi=\Pi_2^s H_{\omega,s}(1-S)\psi$ 
so that for $\Re e(z)>0$

\begin{equation}
\label{eq-reflechalf}
e^{-z{H}_{\omega,s}}\;=\;\Pi_2^s\,e^{-z\tilde{H}_{\omega,s}}\,(1-S_s)\Pi_2^s
\mbox{ . }
\end{equation}

Furthermore let ${\phi}\in C^\infty(\RR)$ be
monotonously increasing, ${\phi}(-\infty)=0$, ${\phi}(\infty)=1$ and
$\mbox{\rm supp}({\phi}')\subset [0,1]$ and set ${\phi}_s(x)={\phi}(s+x)$. 
The following result is 
similar to the discrete case \cite{KRS,EG}.

\begin{thm}
\label{theo-tunnel}
Let $V_\om\in L^\infty(\RR^2)$ and
$F\in C^{k}_c$, $k>6$ and $s<\infty$.
Then $F(H_{\om,s})$ is an integral operator which can be decomposed
as 

$$
F(H_{\om,s})
\;=\;
{\phi}_s\,F(H_\om)+ K_{\om,s}
\mbox{ , }
$$

\noindent where $K_{\om,s}$ form a covariant family of integral
operators the kernels of which satisfy for any $\delta>0$

\begin{equation}
\label{eq-compbound}
|\langle \x|K_{\om,s}|\y\,\rangle|
\;\leq\;
\frac{c_\delta}{1+|x_2+s|^{k-6-\delta}+|y_2+s|^{k-6-\delta}}
\mbox{ . }
\end{equation}

\end{thm}

\noindent
\bew\ Again we set $s=0$, drop the indices
$\omega$ and $s$ and denote the half-plane operator by $\hat{H}$, the
one on the plane by $H$. Furthermore set:

$$
\tilde{H}\;=\;H+P
\mbox{ , }
\qquad
P
\;=\;
(1-\Pi_2)
\left(-2\gamma\,X_2\,D_1+2\gamma^2X^2_2
-V+SVS\right)(1-\Pi_2)
\mbox{ , }
$$

\noindent
One easily verifies the arguments of Section \ref{sec-halfplanekernel}
which imply that also the integral kernel of 
$e^{-z\tilde{H}}$ satisfies the estimates of Proposition
\ref{prop-halfplaneLandau}.
Using (\ref{eq-reflechalf}) and
Duhamel's formula, one gets the following
operator identity on $L^2(\RR\times\RR_+)$:

$$
e^{-z\hat{H}}
\;=\;
\Pi_2 e^{-zH}\Pi_2 +z\int_0^1 dq
\;
\Pi_2 e^{-(1-q)z\tilde{H}}Pe^{-qz{H}}\Pi_2
-
\Pi_2\,e^{-z\tilde{H}}S\Pi_2
\mbox{ . }
$$

\noindent Replacing this into (\ref{eq-fctcalc}), the first term gives
rise to $\Pi_2 F(H)\Pi_2$, which can easily be replaced by 
${\phi}_0 F(H){\phi}_0$ up to
an error satisfying (\ref{eq-compbound}). The third term leads to
$\Pi_2 F(\tilde{H})S\Pi_2$, which according to Proposition 
\ref{prop-Hamilfunction} (holding also for $\tilde{H}$)
can directly be seen to satisfy (\ref{eq-compbound}), even with $\Pi_2$
replaced by $\phi_0$.

\vspace{.2cm}

Now let us consider the second contribution to $e^{-z\hat{H}}$ and
denote it $I(z)$. In order to estimate it, 
it will be used that
the kernel of $e^{-z\tilde{H}}$ satisfies the estimate 
(\ref{eq-complexest}) following from (\ref{eq-diamaghalf}).
Then, using the particular form of $P$ and the estimate
(\ref{eq-compsemi}), one first obtains 

$$
\left|\langle 
\vec{r}\,|Pe^{-q(1+\imath t){H}}|\y\,\rangle\right|
\;\leq\;
c\;
\left(
\frac{|r_2|}{q^{3/2}}\,({1+t^2})^2\,+\,
\frac{r_2^2}{q}\,+\,\frac{|V|}{q}\right)
\,
e^{-\frac{|\vec{r}-\y\,|^2}{8q(1+t^2)}}
\mbox{ . }
$$

\noindent Due to (\ref{eq-complexest}),
one can bound 

\begin{eqnarray}
\left|\langle \x|I(z)
|\y\,\rangle\right|
& \leq &
c\;
\int_0^1 dq \int_{r_2\leq 0}d\vec{r}
\;
\frac{1}{1-q}\;
e^{-\frac{|\x-\vec{r}\,|^2}{8(1-q)(1+t^2)}}
\left(
\frac{|r_2|}{q^{3/2}}\,({1+t^2})^2\,+\,
\frac{r_2^2}{q}\,+\,\frac{|V|}{q}\right)
\,
e^{-\frac{|\vec{r}-\y\,|^2}{8q(1+t^2)}}
\nonumber
\\
& & 
\nonumber
\\
& \leq &
c\,
\int_0^1 dq \int_{0}^\infty\!\! dr_2\;
\frac{q^{1/2}}{(1-q)^{1/2}}\,
\left(
\frac{|r_2|}{q^{3/2}}\,({1+t^2})^2\,+\,
\frac{r_2^2}{q}\,+\,\frac{|V|}{q}\right)
\,
e^{-\frac{|x_2+r_2|^2}{8q(1+t^2)}-
\frac{|r_2+y_2|^2}{8(1-q)(1+t^2)}}
\nonumber
\\
& & 
\nonumber
\\
& \leq &
c\,({1+t^2})^2
e^{-\frac{x_2^2+y_2^2}{8(1+t^2)}}
\nonumber
\mbox{ , }
\end{eqnarray}

\noindent where in the second step 
the integral over $r_1$ was carried out and the
resulting Gaussian factor $e^{-\frac{|x_1-y_1|^2}{8(1+t^2)}}$ simply
bounded by $1$, and the third follows from the estimate
$|x_2-r_2|^2+|r_2-y_2|^2\geq x^2_2+y_2^2+2r_2^2$,
followed by another Gaussian integration (then over all $r_2\in\RR$). 
Just as in the proof of Proposition \ref{prop-Hamilderive}
the desired
bound on the contribution to $F(\hat{H})$ follows.
\eb

\section{Traceclass estimates}
\label{sec-traceclass}


To begin with, $\TV$-traceclass properties on compactly supported smooth
functions of the planar Hamiltonians are examined. 
Proposition \ref{prop-Hamilderive} implies the continuity of the
integral kernel of $D_jF(H)$ so that one obtains the following:

\begin{cor}
\label{cor-Hamilderive}
Let $\partial_j V_\omega\in L^\infty(\RR^2)$, $j=1,2$,
and $F\in C_c^\infty(\RR)$. 
Then $F(H)\in\Aa$ and $D_jF(H)\in\Aa$ are 
$\TV$-traceclass. Their $\TV$-trace can be calculated by {\rm
(\ref{eq-tracecalc})}.
\end{cor}

\vspace{.2cm}

This result allows to transpose the formalism developed in
\cite{BES,SBB} to prove the Kubo formula for tight-binding
Schr\"odinger operators also to continuous
Schr\"odinger operators. For the definition and evaluation of
the edge currents, the following
$\TVh$-traceclass estimates will be important.

\begin{cor}
\label{cor-traceclass}
Let $\Delta\subset\RR$ be a gap of $H_{\omega,\infty}$ 
and $F:\RR\to\RR_+$ be a smooth positive function
supported by $\Delta$. Suppose
$\partial_j V_\omega\in L^\infty(\RR^2)$, $j=1,2$.
Then $F(H)\in\Ta$ and $D_1F(H)\in\Ta$ 
are $\hat{\TV}$-traceclass. Their trace can be calculated by {\rm
(\ref{eq-tracecalc})}.
\end{cor}

\noindent
\bew\ If $\Delta$ is a gap of $H_{\omega,\infty}$, 
then $F(H_{\omega,\infty})=0$ so that the first
term in Theorem \ref{theo-tunnel} vanishes and the second term is
$K=F(H)$. As $F(H)\geq 0$, one calculate
$\TVh(F(H))$ directly using the integral kernels which satisfy
the estimate of Theorem \ref{theo-tunnel}. This immediately
implies that $\TVh(F(H))<\infty$. As $F$ is positive,
$D_1 F(H)=(D_1 F(H)^{1/2})F(H)^{1/2}$. As $F(H)^{1/2}$ is 
$\hat{\TV}$-traceclass by the above argument, so is $D_1 F(H)$.
\eb

\vspace{.2cm}

The next result does not allude to properties of the Hamiltonian, but
rather gives a general property of $\hat{\TV}$-traceclass operators. 
Therefore let $|\nabla_j|$ be new operations on $\Ta$ defined by 
$\langle \x\,|\,(|\nabla_j|A)_\oh|\y\,\rangle=|x_j-y_j|
\langle \x\,|A_\oh|\y\,\rangle$. Whether $|\nabla_j|A\in\Ta$ can, for
example, easily be deduced from (\ref{eq-normbound}) if the integral
kernel of $A$ decays off the diagonal.
Furthermore let us introduce the
function $\Sigma(\x\,)=\mbox{sign}(x_1)$ and denote the associated
multiplication operator also by $\Sigma$.

\begin{prop}
\label{prop-HS}
Suppose that $A\in\Ta$ is $\hat{\TV}$-traceclass and that the integral
kernels are jointly continuous.
Moreover let
$|\nabla_1|A\in\Ta$. Then for any $s<\infty$, the operators
$\left[\Sigma,A_{\omega,s}\right]$ are Hilbert-Schmidt and
the square of their Hilbert-Schmidt norm is $\PP$-integrable.
\end{prop}

\noindent
\bew\
It follows from the hypothesis and the ideal property
that $(|\nabla_1|A^*)A$ is
$\hat{\TV}$-traceclass and therefore

\begin{eqnarray*}
\TVh((|\nabla_1|A)^*\,A)
&=&
\int_\RR ds \int_\Omega d\PP(\omega)
\int d\y\;|y_1| 
\left|\langle \y\,|A_{\omega,s}|0\rangle\right|^2 \\
\end{eqnarray*}

\noindent is finite.
Replacing the identity

$$
|y_1|
\;=\;
\frac{1}{2}\;
\int dx_1\;
\left(1-\Sigma(\y+\x\,)\Sigma(\x\,)\right)
\mbox{ , }
$$

\noindent and using the covariance relation one obtains

\begin{eqnarray*}
\TVh((|\nabla_1|A)^*\,A)
&=&\frac{1}{2}\;\int d\PP(\omega)
\int d\x\int d\y\;(1-\Sigma(\y)\Sigma(\x)) 
\left|\langle \y\,|A_{-\x\cdot\omega,s}|\x\,\rangle\right|^2 \\
&=&\frac{1}{4}\;\int d\PP(\omega)
\int d\x\;
\langle \x\,|[\Sigma,A_{-\x\cdot\omega,s}]^*
[\Sigma,A_{-\x\cdot\omega,s}]|\x\,\rangle \\
&=&\frac{1}{4}\;
\int d\PP(\omega)\;
\Tr\left(
\left|\left[\Sigma,A_{\omega,s}\right]\right|^2
\right)
\mbox{ , }
\end{eqnarray*}

\noindent where in the last step
the integrations were exchanged and the translation invariance of
$\PP$ was used.
This implies the claim because of the weak continuity of
$A_{\omega,s}$ in $\omega$.
\eb

\section{Currents}

By the Heisenberg equations of motion
the current operators are given by 

\begin{equation}
\label{eq-currentop}
J_j
\;=\;
\left.\frac{d}{dt} X_j(t)\right|_{t=0}
\;=\;
\imath[H_{\oh},X_j]
\;=\;
-2D_j
\mbox{ , }
\qquad
j=1,2
\mbox{ . }
\end{equation}

\noindent Accessible in experiment is the expectation value of the
current w.r.t. to a given one-particle density matrix $\rho$.
The current density in the bulk is then calculated in the planar model
using the trace per unit volume $\TV$. 
The following result implies that no
bulk current flows at equilibrium and absence of electric field,
that is, if the density matrix is a function of the Hamiltonian such as
the Fermi-Dirac function. 
This result 
was already given in \cite{BES}, but only with a very sketchy proof.

\begin{prop}
\label{prop-nocurrent}
Let $F\in C_c^{k}(\RR)$ with $k>5$. Then 

$$
\TV(J_j F(H))\;=\;0
\mbox{ . }
$$
\end{prop}

\noindent \bew\ 
Let us begin by noting that $\nabla_j$-invariance of $\TV$ and
Duhamel's formula imply that for $\Re e(z)>0$

$$
0\;=\;\TV(\nabla_j e^{-zH})
\;=\;
z\;\TV
\left(\int^1_0 dq\, e^{-(1-q)zH}
(\nabla_j H) e^{-qzH}\right)
\mbox{ . }
$$

\noindent Since $e^{-qzH}$ is $\TV$-traceclass only for $q>0$,
the integral $\int^1_0=\int^1_{\frac{1}{2}}+
\int^{\frac{1}{2}}_0$ is split. 
This allows to use cyclicity in order to obtain

$$
0\;=\;
z\;\TV
\left(
(\nabla_j H) e^{-zH}\right)
\mbox{ . }
$$

\noindent 
Finally the representation by a
norm convergent Riemann integral (\ref{eq-fctcalc}) 
can be used to conclude
$$
\TV(J_j\,F(H))
\;=\;
-2\int_\RR dt\,\tilde{F}(t)\;
\TV \left(
(\nabla_j H) e^{-(1+\imath t)H}\right)
\;=\;0
\mbox{ , }
$$

\noindent where the trace $\TV$ and the sum defining the Riemann
integral over $t$ could be exchanged because $e^{-(1+\imath t)H}$ is
$\TV$-traceclass for any $t$ due to the results of Section
\ref{sec-kernelplane}. 
\eb 

\vspace{.2cm}

For a system with a boundary in the $1$-direction, 
an edge current flows 
along the infinite boundary of the half-plane. 
However, this current only flows in the vicinity of the boundary so
that the trace per unit volume $\TV$ of $J_1\rho$ 
vanishes. In fact, the physical
current density along the boundary is rather obtained by
taking the trace per unit
volume in the $1$-direction followed by the usual trace in the
$2$-direction, an operation precisely given by $\TVh$. 
Corollary \ref{cor-traceclass} implies that the
following definition of the edge current is also mathematically sound as
long as $F$ is positive and supported by a gap of $H_{\omega,\infty}$:
\begin{equation}
\label{eq-edgecurrentdef}
j^e(F)
\;=\;
\TVh(J_1\,F(H))
\mbox{ . }
\end{equation}
One might erroneously
believe that analogous to Proposition \ref{prop-nocurrent}
one has $\TVh({J}_1F({H}))=0$ at least if $F$ is supported by a
gap of $H$. In fact, the proof of
Proposition \ref{prop-nocurrent} does not carry over because the
semigroup is {\it not} $\TVh$-traceclass. What the (finite)
value of $\TVh({J}_1F({H}))$ is, will be analyzed in the next sections.

\vspace{.2cm}

At this point let us comment on 
what happens if the spectrum of $H_\omega$ does not have a
gap. Then $F(H_{\omega,s})=\phi_s F(H_{\omega,\infty})+K_s$ 
where $\phi_s F(H_{\omega,\infty})$ is
definitely not $\TVh$-traceclass and Theorem \ref{theo-tunnel} implies
that $K_s$ is a boundary operator, 
although it does not directly imply that $K_s$ is moreover
$\TVh$-traceclass because it may not have a definite sign. In order to
make nevertheless sense of the edge current in this situation, one can
regularize the expression and rather define the edge current by

$$
j^e(F)\;=\;
\lim_{S\to\infty}
\,\int^S_{-S}ds
\;\int d\PP(\omega)\;
\langle \vec{0}\,|J_1 F({H}_{\omega,s})|\vec{0}\,\rangle
\mbox{ . }
$$

\noindent Due to Proposition \ref{prop-nocurrent}, one then sees that
the contribution coming from $\phi_s F(H_{\omega,\infty})$ 
vanishes for every finite
$S$. Hence, assuming that the remainder $K_s$ is actually
$\TVh$-traceclass, one then obtains $j^e(F)=\TVh(J_1K_s)<\infty$, hence
a reasonable definition. 

\vspace{.2cm}

\section{Winding numbers}
\label{sec-winding} 

On $\Ta\times\Ta$ consider the densely defined bilinear map 

\begin{equation}
\label{def-1cocyc}
\xi(A,B)
\;=\;
\imath\;
\TVh(A\nabla_1 B)
\mbox{ , }
\end{equation}

\noindent 
If $A$ is $\TVh$-traceclass and $\nabla_1 B\in\Ta$ (or vice versa) then
$(A,B)$ belongs to the domain of definition of $\xi$ denoted ${\cal D}(\xi)$.

\begin{lem}
\label{lem-1cocycle} $\xi$ is a $1$-cocycle, namely it satisfies
whenever $(A,B), (B,C)(C,A)
\in {\cal D}(\xi)$ have jointly continuous integral kernels:

\vspace{.1cm}

\noindent {\rm (i)} Cyclicity: $\xi(A,B)=-\xi(B,A)$.

\vspace{.1cm}

\noindent {\rm (ii)} Closedness under the Hochschild operator:
$\xi(AB,C)-\xi(A,BC)+\xi(CA,B)=0$.

\end{lem}

\noindent
\bew\ This follows from a short algebraic 
calculation using the Leibniz rule
for $\nabla_1$ and the $\nabla_1$-invariance of $\TVh$ holding under
the stated hypothesis.
\eb

\vspace{.2cm}

By general principles \cite{Con} (see also \cite{KRS,KS}), 
$1$-cocycles can be paired with unitaries. The pairing in the present
context stems from a
Fredholm module so that it leads to an index theorem. Let
$\Pi_1$ be the indicator function on the half-space with
positive first coordinate, i.e.\  $\Pi_1=\frac{1}{2}(\Sigma+1)$.
The projection from $L^2(\RR^2)$ onto  $L^2(\RR_+\!\times\!\RR)$
is also denoted by $\Pi_1$. 

\begin{thm}
\label{theo-winding}
Let $\Uu$ be a unitary such that $\Uu-1\in\Ta$ is
$\TVh$-traceclass and has jointly continuous integral
kernel. Furthermore let $\nabla_1 \Uu\in\Ta$ and $|\nabla_1\!|\, \Uu\in\Ta$.
Then for fixed $s<\infty$ and $\om\in\Omega$, $\Pi_1\,\Uu_{\oh}\Pi_1$
is a Fredholm operator on $L^2(\RR_+\!\times\RR)$. 
If $\vec{\xi}\in\RR^2\mapsto \Uu_{\vec{\xi}\cdot\om,s}$ is moreover 
norm-continuous,
the corresponding index $\mbox{\rm Ind}$
is $\PP$-almost surely independent of 
$\omega$, always independent of $s$, and given by
$$
\mbox{\rm Ind}\;=\;-\,\xi(\Uu^*-1,\Uu)
\mbox{ . }
$$
\end{thm}

\noindent
\bew\ By Proposition \ref{prop-HS}, the conditions imply that
$[\Sigma,\Uu_{\oh}]$ is Hilbert-Schmidt. From 
the algebraic identity:

\begin{equation}
\label{eq-algid}
\Pi_1 A_{\oh}B_{\oh}\Pi_1-
\Pi_1 A_{\oh}\Pi_1 B_{\oh}\Pi_1
\;=\; - \frac{1}{4}\;\Pi_1\,
\left[\Sigma,A_{\oh}\right]\left[\Sigma,B_{\oh}\right]
\mbox{ , }
\qquad
A,B\in\Ta
\mbox{ , }
\end{equation}

\noindent follows that $\Pi_1-\Pi_1 \Uu_{\oh}\Pi_1 \Uu_{\oh}^*\Pi_1$
and $\Pi_1-\Pi_1 \Uu_{\oh}^*\Pi_1 \Uu_{\oh}\Pi_1$
are traceclass. 
By Fedosov's formula ({\sl e.g.\ }
\cite{Con,BES,KRS}),  $\Pi_1 \Uu_{\oh}\Pi_1$ is a Fredholm operator
on $L^2(\RR_+\!\times\!\RR)$ whose index 
is given by

$$
\mbox{Ind}_\oh
\;=\;
\Tr\left(\Pi_1-\Pi_1\, \Uu_{\oh}^*\Pi_1\, \Uu_{\oh}\Pi_1\right)-
\Tr\left(\Pi_1-\Pi_1\, \Uu_{\oh}\Pi_1\, \Uu_{\oh}^*\Pi_1\right)
\mbox{ . }
$$

\vspace{.2cm}

By hypothesis,
$\vec{\xi}\in\RR^2\mapsto \Pi_1\Uu_{\vec{\xi}\cdot\om,s}\Pi_1$ is a
norm-continuous family of Fredholm operators so
that by homotopy invariance and ergodicity of $\PP$ their Fredholm index
is $\PP$-almost surely constant. The identity
$\Pi_1\, \Uu_{\omega,s+\xi_2}\Pi_1=U(0,\xi_2)^*
\Pi_1\, \Uu_{(0,\xi_2)\cdot\omega,s}\Pi_1
U(0,\xi_2)$ implies that $s\in\Real\mapsto \Pi_1\,
\Uu_{\omega,s}\Pi_1$ is 
norm continuous and therefore $\mbox{Ind}_{\om,s}$ constant in $s$. 

\vspace{.2cm}

The almost sure index $\mbox{Ind}$ is, for any $s\in\RR$, given by

$$
\mbox{Ind} \;=\; \int d\PP(\om) \;
\mbox{Ind}_{\om,s}
\;=\; -\;\eta_s(\Uu^*-1,\Uu)
\mbox{ , }
$$

\noindent where $\eta_s(A,B)=\int d\PP(\omega)\,
\eta_{\omega,s}(A,B)$ with  
$$
\eta_\oh(A,B)
\;= \;
\Tr\left(\Pi_1 B_{\oh}A_{\oh}\Pi_1-
\Pi_1 B_{\oh}\Pi_1 A_{\oh}\Pi_1\right) 
-
\Tr\left(\Pi_1 A_{\oh}B_{\oh}\Pi_1-
\Pi_1 A_{\oh}\Pi_1 B_{\oh}\Pi_1\right) 
\mbox{ . }
$$

\vspace{.2cm}

\noindent Introduce next the $1$-cocycle $\zeta_s$

$$
\zeta_s(A,B)
\;=\;
\int d\PP(\omega)\;
\zeta_{\omega,s}(A,B)
\mbox{ , }
\qquad
\zeta_\oh(A,B)
\;=\;
\frac{1}{4}\,
\Tr\left(\Sigma\left[\Sigma,A_{\omega,s}\right]
\left[\Sigma,B_{\omega,s}\right]\right)
\mbox{ . }
$$

\noindent Then $\eta_\oh(A,B)=\zeta_\oh(A,B)$ because of identity
(\ref{eq-algid}) and the cyclicity property of $\zeta_\oh$.

\vspace{.2cm}

\noindent 
Using the invariance of
$\PP$ as well as the identity 

$$
\int dy_1
\;\Sigma(\y)\;\left(\Sigma(\y)-\Sigma(\y+\x)\right)^2
\;=\;
-4\,x_1
\mbox{ , }
$$

\noindent one can verify as in the proof of Proposition~\ref{prop-HS}
that $\xi(A,B)=\zeta_s(A,B)$ for all finite $s$.
\eb

\vspace{.2cm}

The above calculations follow closely
\cite[Sec.~4.2,~4.3]{KRS}. 
However, there is one crucial
difference. The invariance of the index in the $2$-direction holds for
all unitaries, while in the discrete case it was only true for unitaries
in the image of the exponential map \cite[Prop.~4.10]{KRS}. The reason
is that the exponential map is an isomorphism in the continuous case,
namely it is Connes' Thom isomorphism \cite{Con81}. Further
explanations will be given in \cite{KS}.

\vspace{.2cm}

Note that $\Uu-1$ being
$\TVh$-traceclass implies that also $\Uu^k-1$ is
$\TVh$-traceclass for any $k\in\ZZ$. In fact, this follows from 
$\Uu^k-1=(\Uu-1)\sum_{l=0}^{k-1}\Uu^l$ and the fact that traceclass
operators form an ideal. It is then elementary to verify that under
the assumptions of the theorem

\begin{equation}
\label{def-windprop}
\eta_s((\Uu^*)^k-1,\Uu^k)\;=\;k\,\eta_s(\Uu^*-1,\Uu)
\mbox{ . }
\end{equation}


\section{Quantization of edge currents}
\label{sec-edgecond} 

Let $\Delta=[E',E'']$ be in a gap of the spectrum of
$H_{\omega,\infty}$. Let $G\in C^\infty(\RR)$ be a
monotonously decreasing function with $G(-\infty)=1$, $G(\infty)=0$, and
$\mbox{\rm supp}(G')\subset \Delta\backslash G^{-1}(\frac{1}{2})$. The
support of a function is closed by definition
and hence all derivatives of $G$ vanish on the pre-image $G^{-1}(\frac{1}{2})$.
Define via functional calculus the following unitary
operator 
\begin{equation}
\label{def-unitary}
{\cal U}(\Delta)
\;=\;
\exp (-2\pi \imath \,G({H}))
\mbox{ . }
\end{equation}


\begin{thm} \label{thm3}
Suppose $\partial_j V_\omega\in L^\infty(\RR^2)$ for $j=1,2$.
Let $\Delta$ be in a gap of the spectrum of
$H_{\omega,\infty}$. 
Then $J_1G'({H})\in\Ta$ and ${\cal
U}(\Delta)-1\in\Ta$ are both
$\TVh$-traceclass, $\nabla_1{\cal U}(\Delta)\in\Ta$ and
for $\PP$-almost all $\omega$ and all $s\in\RR$,
\begin{equation}
\label{eq-curcalc}
-\,2\pi\;
\TVh(J_1G'({H}))
\;=\;
\imath\,\TVh(({\cal U}(\Delta)^*-1)\,{\nabla}_1{\cal U}(\Delta)) 
\;=\;
\mbox{{\rm Ind}}(\Pi_1\,{\cal U}_{\omega,s}(\Delta)\Pi_1)
\mbox{ . }
\end{equation}

\end{thm}

\noindent
\bew\ First the assumptions of
Theorem~\ref{theo-winding} are established. 
As the function $G$ is monotonously decreasing, $G'$ is a
negative smooth function supported by $\Delta$ so that
Corollary~\ref{cor-traceclass} implies that $J_1G'({H})$ is
$\TVh$-traceclass. 
To prove the $\TVh$-traceclass property of 
${\cal U}(\Delta)-1$, let us write it as the linear combination of three
positive and smooth functions of $H$.
Indeed, $E\mapsto \sin_\pm(2\pi \,G(E))$ and
$E\mapsto \cos(2\pi\, G(E))-1$ (where $g_\pm$ denotes 
the positive and negative parts of a real
function $g$) are of this type since $\sin_\pm(2\pi \,G(E))$ vanishes
with all its derivatives at $E\in G^{-1}(\frac{1}{2})$. That
$\nabla_1{\cal U}(\Delta)\in\Ta$ and 
$|\nabla_1|{\cal U}(\Delta)\in\Ta$ follows directly from 
Proposition~\ref{prop-Hamilfunction} for the half-plane operators
combined with (\ref{eq-normbound}).
Finally, by Duhamel's formula

$$
\vec{\xi}\in\RR^2\;\mapsto\;
e^{-t H_{\vec{\xi}\cdot \omega,s}}
\;=\;
\int^1_0 dq\;
e^{-(1-q)t H_{L,s}}
\,V_\omega(.-\vec{\xi})\,
e^{-qt H_{\vec{\xi}\cdot \omega,s}}
\mbox{ , }
$$

\noindent so that the continuity of the potential implies the
norm-continuity of the semigroups and via the norm-convergent
functional calculus (\ref{eq-fctcalc}) also of
$\vec{\xi}\in\RR^2\;\mapsto\;{\cal U}(\Delta)_{\vec{\xi}\cdot
\omega,s}$. In conclusion,
the conditions of Theorem \ref{theo-winding} are verified and only the
first equality in (\ref{eq-curcalc}) remains to be shown.

\vspace{.2cm}

For that 
express ${\mathcal U}(\Delta)$ as exponential series and use the Leibniz
rule to obtain

$$
\TVh(({\cal U}(\Delta)^*-1)\,{\nabla}_1{\cal U}(\Delta)) 
\;=\; 
\sum_{m=0}^\infty
\frac{(-2\pi\imath)^m}{m!}
\;\sum_{l=0}^{m-1}\;
\TVh
\left(({\cal U}(\Delta)^*-1)\,G({H})^l\,
{\nabla}_1 G({H})\,G({H})^{m-l-1}
\right)
\mbox{ , }
$$

\noindent where the trace and the infinite sum could be exchanged
because of the traceclass properties (note that also
$\nabla_1 G({H})\in\Ta)$. Due to cyclicity and the fact that 
$[{\cal U}(\Delta),G({H})]=0$, each summand is now equal to 
$\TVh(({\cal U}(\Delta)^*-1)\,G({H})^{m-1}\,
{\nabla}_1G({H}))$. Exchanging again sum and trace and
summing the exponential up again, one gets

$$
\TVh(({\cal U}(\Delta)^*-1)\,{\nabla}_1{\cal U}(\Delta)) 
\;=\;
-2\pi\imath\;\TVh
\left((1-{\cal U}(\Delta))\,{\nabla}_1G({H})\right)
\mbox{ . }
$$

\noindent Repeating the same argument for $\Uu(\Delta)^k=\exp(-2\pi
\imath \,k\,G({H}))$ where
$k\in\ZZ$ and using (\ref{def-windprop}) more generally
implies that, for  $k\neq 0$,

$$
\TVh
\left((1-{\cal U}(\Delta))\,{\nabla}_1G({H})\right)
\;=\;\TVh
\left((1-{\cal U}(\Delta)^k)\,{\nabla}_1G({H})\right)
\mbox{ . }
$$

\noindent 
Writing 
$G(E)=\int dt\,\tilde{G}(t)\,e^{-E(1+\imath t)}$ as in 
(\ref{eq-fctcalc}), the above r.h.s.\ is, for $k\neq 0$, equal to
$$
-2\pi\imath\;
\int dt\,\tilde{G}(t)\,(1+\imath t)\;
\int^1_0dq\;
\TVh
\left((1-{\cal U}(\Delta)^k)\,
e^{-(1-q)(1+\imath t){H}}
({\nabla}_1 H)
e^{-q(1+\imath t){H}}
\right)
\mbox{ . }
$$

\noindent 
The integral over $t$ is a norm convergent
Riemann integral. One therefore finds using
$G'(E)=-\int dt\,(1+\imath t) \,\tilde{G}(t)\,e^{-E(1+\imath t)}$,
for $k\neq 0$,
$$
\TVh(({\cal U}^*(\Delta)-1)\,{\nabla}_1{\cal U}(\Delta)) 
\; = \;
2\pi\imath\;
\TVh
\left(({\cal U}(\Delta)^k-1)\,
({\nabla}_1H)
\,G'({H})
\right)
\mbox{ , }
$$

\noindent while, for $k=0$, the r.h.s. vanishes.

\vspace{.2cm}

To conclude, let $\phi:[0,1]\to \RR$ be a differentiable function
vanishing at the boundary points $0$ and $1$.
Let its Fourier coefficients be denoted by 
$a_k=\int^1_0 dx \,e^{-2\pi\imath k x}\phi(x)$. Then 
$\sum_k a_k e^{2\pi\imath k x}=\phi(x)$ and, in particular, $\sum_k a_k=0$.
Hence

\begin{eqnarray}
\left(\sum_{k\neq 0} a_k\right) \,
\TVh(({\cal U}^*(\Delta)-1)\,{\nabla}_1{\cal U}(\Delta)) 
& = &
2\pi\imath\;
\sum_{k} a_k\;
\TVh
\left(({\cal U}(\Delta)^k-1)\,
({\nabla}_1H)
\,G'({H})
\right)
\nonumber
\\
& = &
2\pi\imath\;\TVh
(G'({H})\,\phi(G(H))\,({\nabla}_1H))
\mbox{ . }
\nonumber
\end{eqnarray}

\noindent Let now $\phi$ converge to the indicator function of
$[0,1]$. Then $a_0\to 1$ and $\sum_{k\neq 0} a_k\to -1$, 
while $G'({H})\phi(G(H))\to 
G'({H})$ (the Gibbs phenomenon is
damped). As $J_1=-\nabla_1 H$, this concludes the proof.
\eb

\vspace{.2cm}

\section{Link to bulk Hall conductivity}
\label{sec-bulkcond}

\vspace{.2cm}

The Chern
character is a trilinear form 
$\mbox{ch}$ defined by

\begin{equation}
\label{eq-Chern}
\mbox{ch}(A,B,C)
\;=\;2\pi\imath\,
\TV(A((\nabla_1B)(\nabla_2C)-(\nabla_2B)(\nabla_1C)))
\mbox{ , }
\qquad
A,B,C\in\Aa
\mbox{ , }
\end{equation}

\noindent 
as long as the r.h.s. is well-defined. 
The following theorem is well-known \cite{Kun,Bel,ASS}.

\begin{thm} 
\label{theo-bulkHall}
Let $P\in\Aa$ be a projection with integral kernels 
satisfying $\PP$-a.s. for some $\delta>0$

\begin{equation}
\label{eq-loccond}
\left|\langle \x|P_\omega|\y\,\rangle\right|
\;\leq\;
\frac{c_\delta}{1+|\x-\y\,|^{2+\delta}}
\mbox{ . }
\end{equation}

\noindent Then $\mbox{ch}(P,P,P)$ is well-defined and equal to an integer
given as the index of a Fredholm operator.
\end{thm}

The importance of this result stems from the fact that  
the bulk Hall conductivity of a gas of
independent electrons described by $H$ 
at zero-temperature, zero dissipation and
with chemical potential $\mu$ is given by
\cite{AS,Kun,Bel,NB,ASS,BES,AG}

$$
\sigma^\perp_b(\mu)
\;=\;
\frac{q^2}{h}\;\mbox{ch}(P_\mu,P_\mu,P_\mu)
\mbox{ , }
$$

\noindent where $P_\mu=\chi_{(-\infty,\mu]}(H)$ 
is the family of associated Fermi projections. This fact can be
deduced from Kubo's formula \cite{BES,AG} or the adiabatic Laughlin {\sl
Gedankenexperiment}  \cite{ASS}. 

\vspace{.2cm}

As discussed in great detail in \cite{BES,AG} in the discrete setting,
(\ref{eq-loccond}) is a dynamical localization condition on the spectral region
in the vicinity of the Fermi level. For the purposes of the present article,
however, we restrict ourselves to the situation where
the Fermi level $\mu$ is in a gap of the spectrum of $H$. Then
$P_\mu$, defined with a characteristic function, can also be written
as a smooth function of $H$ for which the estimate (\ref{eq-loccond})
holds by Proposition \ref{prop-Hamilfunction}. By homotopy of a
Fredholm index, one then deduces:

\begin{cor} 
\label{coro-bulkHall}
Let the interval $\Delta$ be a gap of the spectrum of
$H_{\omega,\infty}$.  Then
$\mu\in\Delta\mapsto \sigma_\perp^b(\mu)$ is constant and equal to an integer
multiple of $\frac{q^2}{h}$.
\end{cor}

The following result, analogous to the discrete case
\cite{SKR,KRS,EG}, albeit based on Connes' Thom
isomorphism and its dual in cyclic cohomology \cite{Con81,ENN}
instead of the Pimsner-Voiculescu sequence and its dual,
will be proven in \cite{KS}:

\begin{thm} 
\label{theo-link}
Let the interval $\Delta$ be a gap of
$H_{\omega,\infty}$.  Then
$\sigma_\perp^b(\mu)=\mbox{{\rm Ind}}(\Pi\,\Uu_\oh(\Delta)\Pi)$ for
$\mu\in\Delta$. 
\end{thm}

\vspace{.2cm}

\noindent {\bf Acknowledgment:} This work was supported by the
SFB 288 ``Differentialgeometrie und Quantenphysik''. 


\end{document}